\documentclass[conference,twocolumn]{IEEEtran}
\usepackage{fixltx2e}
\usepackage[latin1]{inputenc}
\usepackage{amsmath}
\usepackage{amsfonts}
\usepackage{amssymb}
\usepackage{slashbox}
\usepackage{pict2e}
\usepackage{epsfig}
\usepackage{psfrag}
\usepackage{array}
\usepackage{verbatim}
\usepackage{threeparttable} 
\usepackage{algorithm}
\usepackage{algorithmic}
\usepackage{color}

\renewcommand{\mag}{\text{Mag}}

\newcommand{\bec}{}

\begin{document}
\IEEEoverridecommandlockouts
\title{Deterministic Differential Search  Algorithm for Distributed Sensor/Relay Networks}
\author{\IEEEauthorblockN{Po-Chun~Fu, Pei-Rong~Li, Li-Ming~Wei, Chang-Lin~Chen, and Che Lin$^\dagger$}
\thanks{$^\dagger$Che Lin is the corresponding author. This work is supported by National Science Council, Taiwan (R.O.C.), under Grant NSC102-2221-E-007-024.}
\IEEEauthorblockA{Institute of Communication Engineering \& Department of Electrical Engineering\\
National Tsing Hua University, Hsinchu, Taiwan 30013 \\
}E-mail: s9920101@m99.nthu.edu.tw,~s9961151@m99.nthu.edu.tw,~markwei27990309@yahoo.com.tw\\
~vip410246@gmail.com,~clin@ee.nthu.edu.tw}

\maketitle
\noindent 

\section{Introduction}\label{sec:intro}

For distributed sensor/relay networks, high reliability and power efficiency are often required. However, several implementation issues arise in practice. One such problem is that all the distributed transmitters have limited power supply since the power source of the transmitters cannot be recharged continually. To resolve this, distributed beamforming has been proposed as a viable solution where all distributed transmitters seek to align in phase at the receiver end. However, it is difficult to implement such transmit beamforming in a distributed fashion in practice since perfect channel state information (CSI) need to be made available at all distributed transmitters, requiring tremendous overhead to feed back CSI from the receiver to all distributed transmitters. 

In literature, the efforts of designing efficient distributed phase alignment algorithms can be categorized into two class: deterministic phase adjustment or random phase perturbation algorithms. In \cite{Mudumbai2006,Mudumbai2010}, Mudumbai et al proposed and provided initial analysis for an adaptive distributed beamforming which requires only a single bit feedback. In \cite{Lin2006,Chen2010}, a general set of adaptive distributed beamforming algorithms was reformulated as a local random search algorithm and analyzed. A bio-inspired robust adaptive random search algorithm was proposed and analyzed in \cite{Tseng2011}.

On the other hand, several deterministic distributed beamforming algorithms were also proposed. For example, Thibault et al introduced  a deterministic algorithm with individual power constraint \cite{Thibault2010}. For amplify-and-forward wireless relay networks, an algorithm using additive deterministic perturbations  was presented in \cite{Fertl2008}. 
Fertl et al further investigated a multiplicative deterministic perturbations for distributed beamforming under a total power constraint \cite{Fertl2009}.

In this paper, we propose a novel algorithm that belongs to the category of deterministic phase adjustment algorithm: the Deterministic Differential Search Algorithm (DDSA), where the differences between the measured received signal strength (RSS) are utilized judiciously to help us predict the deterministic phase adjustment done at distributed transmitters. Numerical simulations demonstrate rapid convergence to a pre-determined threshold. 

\section{Results and Discussions}
In this study, we consider a wireless sensor/relay network consisting of distributed transmitters where  a common message $ s \in C$ is to be conveyed to the receiver end. Let $N_s$ be the total number of all distributed transmitters each with a power constraint of $E[|s|^2]$. Each transmitter and the receiver is assumed to be equipped with single antenna. We assume that the channel between them is frequency flat and slow fading. In addition, there exists a one-bit, error-free feedback link from the receiver to all distributed transmitters.

According to system specification, the discrete-time, complex baseband model is given by 
\begin{equation}\label{eq:model}
\begin{aligned}
&y[n] = \sum_{i=1}^{N_s} h_i g_i[n] s + w[n] = \sum_{i=1}^{N_s} a_i b_i[n] e^{j(\phi_i + \psi_i[n])} s + w[n]
\end{aligned}
\end{equation}
where $y[n]$ denotes the received signal, $h_{i} = a_i e^{j \phi_i}$ correspond to the channel coefficients, $g_{i}[n] = b_{i}[n]e^{j \psi_{i}[n]} $ correspond to the beamforming coefficients, and $w[n]$ is the additive white Gaussian noise.

To simplify notations and deal with the problem of distributed phase alignment easily, we let $\theta_i[n]=\phi_i+\psi_i[n]$ be the total phase of received signal from i-th transmitter during the n-th transmission time slot. Also, we denote $s=\sqrt{P}$ and impose fixed power constraint $b_i=1$ among transmitter. Therefore, the received signal can be expressed as $\sqrt{P}\sum_{i=1}^{N_s}a_ie^{j\theta _i [n]}$. Finally, we assume that the strength of the composite signal from receiever can be perfectly estimated at the destination and the {received signal strength (RSS)} function can be described by 
\begin{equation}\label{eq:mag}
\mag(\theta_1[n],\cdots,\theta_{N_s}[n]) = \sqrt{P} \left | \sum_{i=1}^{N_s} a_i e^{j\theta_i[n]} \right |
\end{equation}

Note that the primary goal here is to maximize the RSS function given in (\ref{eq:mag}) for distributed beamforming schemes to exploit the potential power gain efficiently. Furthermore, it is clear that if we want to reach the global maximum, all phase should be completely alignment, i.e, $\theta_1[n]=\theta_2[n]=\cdots =\theta_{N_s}[n]$. 

The proposed DDSA algorithm undergoes $N_s$ rounds of phase evaluations corresponding to the $N_s$ distributed transmitters.
During the $i$-th round, only the $i$-th transmitter transmits for $2$ iterations. 
The $i$-th transmitter alters its phase by iterating through the elements in the set $\Theta_i$ given by: 
\[
\Theta_i = \left \{ \theta_i[0] + \alpha  j,~ j = 0, 1, 2 \right \} 
\]
where $\theta_i[0]$ is the initial phase and {$\alpha=2\pi/3$}. The corresponding RSS measured at the receiver can hence be expressed as
\[
M_j=\left \{ \mag(\theta_i[0] + \alpha  j),~ j = 0, 1, 2\right \} 
\]


Note that the RSS function in (\ref{eq:mag}) can be rewritten as 
\begin{eqnarray}
&\mag(\theta_1[n],\cdots,\theta_{N_s}[n]) = \sqrt{P} \left | \sum_{i=1}^{N_s} a_i e^{j\theta_i[n]} \right | \\
& = \sqrt{P} \left | \sum_{\substack{k=1 \\ k\neq i}}^{N_s} a_k e^{j\theta_k[n]} + a_i e^{j\theta_i[n]} \right | \\
& = \sqrt{P} \left | {r}_i + {c}_i  \right | \\
& = \sqrt{P} \sqrt{\left(\left| r_i \right| + \left|  c_i\right|\cos\beta \right)^2 + \left( \left|  c_i\right|\sin\beta \right)^2}
\end{eqnarray}
where ${r}_i=\sum_{ k\neq i} a_k e^{j\theta_k[n]}$, ${c}_i=a_i e^{j\theta_i[n]}$ and $\beta$ is the phase angle between complex numbers $r_i$ and $c_i$. From this, we have
\begin{equation} \label{eq:Mj}
\begin{aligned}
M_j^2 &= \left(~ |r_i |+ |c_i |\cos\left(\beta+\dfrac{2\pi}{3}j\right)\right)^2\\
&~~~~~+\left(|c_i |\sin\left(\beta+\dfrac{2\pi}{3}j\right)\right)^2\\
&=  |r_i |^2+ |c_i |^2+2 |r_i | |c_i |\cos\left(\beta+\dfrac{2\pi}{3}j\right), \quad\forall j = 0, 1 ,2
\end{aligned}
\end{equation} 

Note that \eqref{eq:Mj} describes $3$ linearly independent equations with $3$ independent variable $|r_i|$, 
$|c_i|$ and $\beta$. Therefore, we can obtain $|r_i|$, $|c_i|$ and $\beta$ by solving \eqref{eq:Mj}. Once $\beta$ is obtained, we can predict and adjust the phase at the $i$th distributed transmitter accordingly. Ideally, one can feed back $\beta$ directly to the $i$th distributed transmitter to achieve perfect phase alignment from the viewpoint of the $i$th transmitter. However, if there is only limited bandwidth for the reverse feedback link, proper quantization is necessary. For example, if $3$ bits are available for the reverse feedback link, $\beta$ can be quantized as $\left\{ 0, \pi/4, \pi/2, 3\pi/4, \pi, 5\pi/4, 3\pi/2, 7\pi/4\right\}$. With this feedback information, the $i$th transmitter can decide whether to subtract either one of $\left\{ 0, \pi/4, \pi/2, 3\pi/4, \pi, 5\pi/4, 3\pi/2, 7\pi/4\right\}$ to achieve a higher RSS function.

\begin{figure}[t]
\includegraphics[width=3.5in]{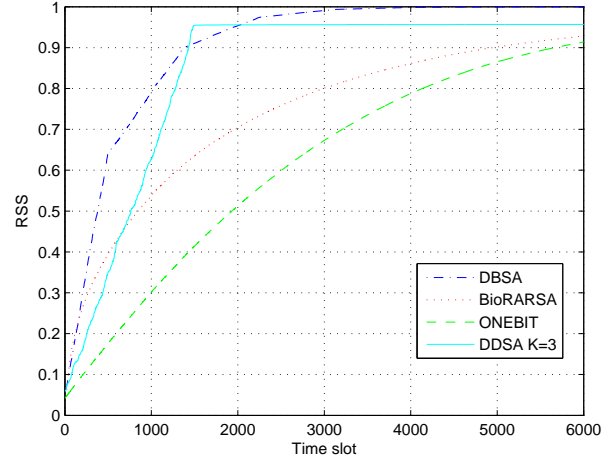}
\caption{Comparison of the convergence behavior between DDSA and other schemes, where $K$ denotes the number of available limited feedback bits.}\label{comparison}
\end{figure}

As a numerical example to demonstrate the rapid convergence of our proposed DDSA, we assume that there are $3$ bits available from the reverse feedback link. Furthermore,  we set the transmitted symbol power to be $\sqrt P= 1$. In our simulation, all channel realizations are assumed to be zero-mean, unit variance  i.i.d Rayleigh flat fading. All simulations are obtained with the same number of distributed transmitters $N_s =  500$. In Figure~\ref{comparison}, we compare the convergence behavior between DDSA and the following schemes: a) one-bit scheme proposed in \cite{Mudumbai2006}, b) BioRASA proposed in \cite{Tseng2011}, and c) DBSA proposed in \cite{DBSA}. We clearly observe that our proposed DDSA exhibit superior convergence behavior and is the first to reach {RSS=0.95}. Therefore, our proposed DDSA presents a efficient and simple alternative to existing adaptive distributed beamforming algorithms. It is important to note that DDSA can also be easily extended to the case where more limited feedback information is available. This is one of our future extensions.



\bibliographystyle{IEEEtran}
\bibliography{ref}
\end{document}